\documentclass[a4paper,11pt]{article}
\pdfoutput=1
\usepackage{jinstpub}
\usepackage{lineno}
\usepackage[utf8]{inputenc}

\title{Development of a single-photon imaging detector with pixelated anode and integrated digital read-out}
\author[a]{ J. A. Alozy}
\author[b]{N. V. Biesuz,\note{Corresponding author.}}
\author[a]{M. Campbell}
\author[c,b]{V. Cavallini}
\author[b]{A. Cotta Ramusino}
\author[c,b]{M. Fiorini}
\author[c,b]{M. Guarise}
\author[a]{X. Llopart Cudie}

\affiliation[a]{CERN, Geneva, Switzerland}
\affiliation[b]{Istituto Nazionale di Fisica Nucleare sezione di Ferrara, Ferrara, Italy}
\affiliation[c]{Università di Ferrara, Ferrara, Italy}

\emailAdd{biesuz@fe.infn.it}

\abstract{
We present the development of a single-photon detector and the connected read-out electronics.
This `hybrid' detector is based on a vacuum tube, transmission photocathode, microchannel plate and a pixelated CMOS read-out anode encapsulating the analog and digital-front end electronics.
This assembly will be capable of detecting up to $10^9$ photons per second with simultaneous measurement of position and time.
    
The pixelated read-out anode used is based on the Timepix4 ASIC ($65~\mathrm{nm}$ CMOS technology) designed in the framework of the Medipix4 collaboration.
This ASIC is an array of $512\times448$ pixels distributed on a $55~\mathrm{\mu m}$ square pitch, with a sensitive area of $\sim 7~\mathrm{cm}^2$. 
It features $50$-$70~\mathrm{e^{-}}$ equivalent noise charge, a maximum rate of $2.5~\mathrm{Ghits/s}$, and allows to time-stamp the leading-edge time and to measure the Time-over-Threshold (\textit{ToT}) for each pixel.
The pixel-cluster position combined with its ToT information will allow to reach $5$-$10~\mathrm{\mu m}$ position resolution.
This information can also be used to correct for the leading-edge time-walk achieving a timing resolution of the order of $10~\mathrm{ps}$.

The detector will be highly compact thanks to the encapsulated front-end electronics allowing local data processing and digitization.
An FPGA-based data acquisition board, placed far from the detector, will receive the detector hits using $16$ electro-optical links operated at $10.24~\mathrm{Gbps}$. 
The data acquisition board will decode the information and store the relevant data in a server for offline analysis.

These performance will allow significant advances in particle physics, life sciences, quantum optics or other emerging fields where the detection of single photons with excellent timing and position resolutions are simultaneously required.
}

\keywords{Hybrid detectors, Photon detectors for UV, visible and IR photons (vacuum), Data acquisition concepts, Front-end electronics for detector read-out}

\arxivnumber{} 

\proceeding{12$^{\text{th}}$ International Conference on Position Sensitive Detectors \\
  12 - 17 September 2021\\
  University of Birmingham, Birmingham, UK}

\date{\today}

\newcommand{\pscalare}{\mathbin{\vcenter{\hbox{$\scriptscriptstyle\bullet$}}}}

\begin{document}

\maketitle
\flushbottom

\section{Introduction}
\label{sec:intro}

Detectors sensitive to a single photon are fundamental in many fields, like research in physics, life sciences or quantum computing. 
Often these applications require excellent timing and spatial resolution.
One example of such applications is high energy physics, indeed due to the ever increasing interaction rate optimal spatial and time resolution are mandatory to distinguish between particles belonging to a single event.
Silicon-based detectors can reach timing resolutions of tens of picoseconds together with spatial resolution of few microns allowing for four-dimensional tracking of charged particles.
Instead, no device with such capabilities is available for single photons.
On the other hand, devices sensitive to single photons over a large active area with tenth of pico-seconds timing resolution are fundamental to allow the use of Cherenkov detectors for particle identification in high pile-up environments, like the one foreseen for the High-Luminosity LHC.

We present the design of a state-of-the-art detector~\cite{Fiorini2018} aimed at the detection of single-photons over a large sensitive area of roughly $7~\mathrm{cm^2}$.
This 'hybrid' detector is based on a vacuum tube, transmission photocatode, micro channel plate stack and a pixelated CMOS read-out anode.
The detector will combine an excellent position resolution ($5-10~\mathrm{\mu m}$) with optimal timing performance ( timing resolution  of the order few tens of picoseconds).
The detector is required to have an high detection efficiency and low dark-rate counts.
The combination of  micro channel plate stack and a pixelated CMOS read-out anode will allow to sustain an high instantaneous rates ($> 10^8~\mathrm{Hits/cm^2/s}$).

\section{Detector Assembly}
\label{sec:assembly}

\begin{figure}
    \centering
    \includegraphics[width=0.5\textwidth]{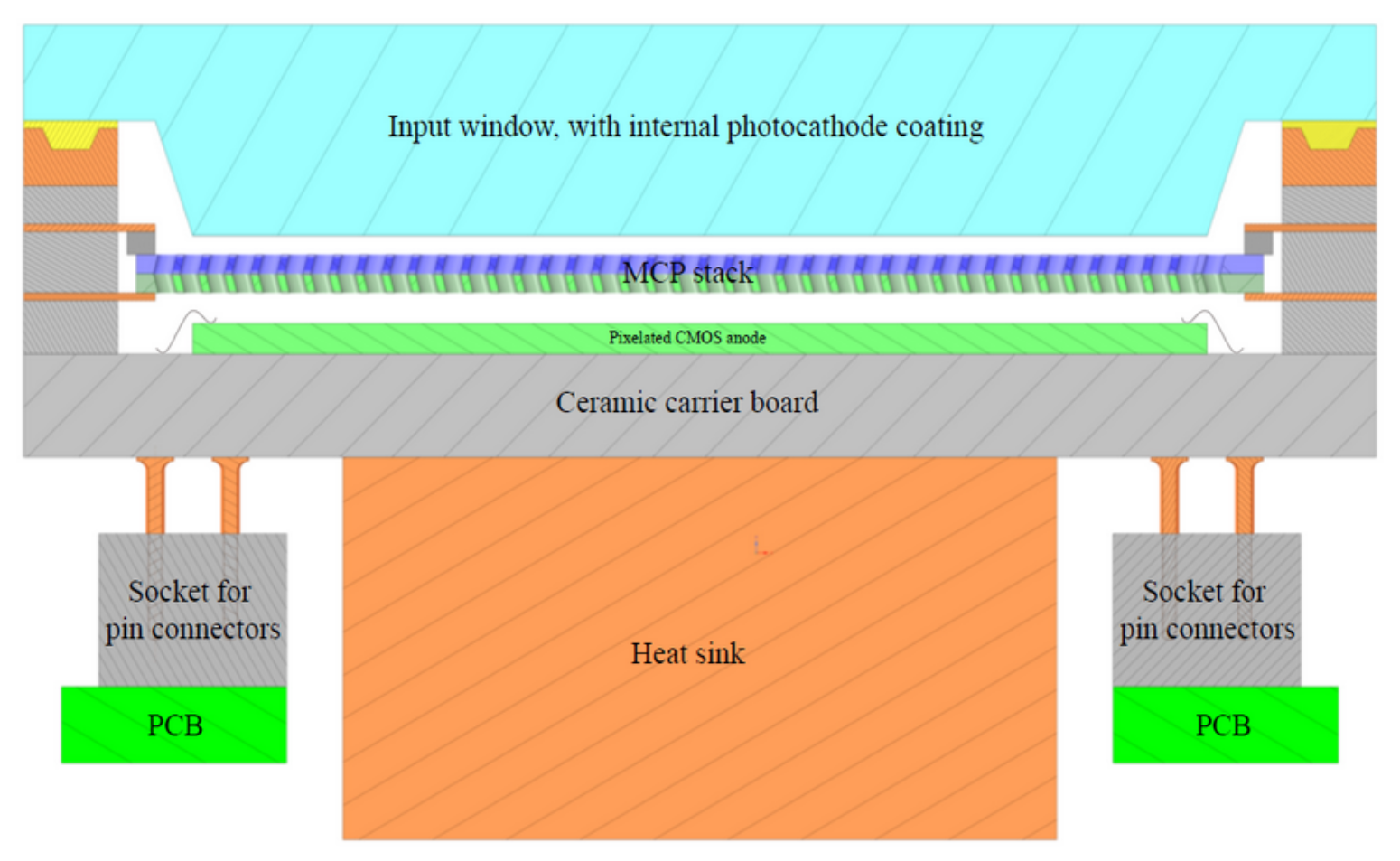}
    \caption{Cutaway schematic view of the detector assembly.}
    \label{fig:assembly}
\end{figure}

A cutaway schematic view of the detector assembly is shown in Figure~\ref{fig:assembly}. 
The detector is based on a vacuum tube with a vacuum level of less than $10^{-10}~\mathrm{mbar}$.
The assembly and in particular the wire-bonding mechanism of the integrated anode is optimized to minimize the distance between components.

The top part of the vacuum tube is composed by a transparent entrance window with an internal high quantum-efficiency photocatode coating.
A drop through window is used to minimize the distance between the photocatode and the multiplication apparatus thus increasing spatial resolution through proximity focusing.
The use of suitable materials for the internal photocatode coating, like bi-alkali or multi-alkali photocatodes, will allow to achieve high quantum efficiencies in the blue-green region of the visible spectrum while limiting the dark count rate to $<10^2~\mathrm{Hz/cm^2}$ at $~300~\mathrm{k}$.
The design of the tube is maintained flexible enough to allow for the use of different photocatodes to increase the quantum-efficiency or to allow for the use of a suitable chiller to decrease the dark count rate.

Electrons emitted by the photocatode will be collected and multiplied by a Micro-Channel Plate (\textit{MCP}) stack composed by two MCPs in chevron configuration.
Each MCP has a pore size of roughly $5~\mathrm{\mu m}$.
We plan to use MCPs treated with a atomic layer deposition to increase the tube lifetime to $20~\mathrm{C/cm^2}$ integrated anode charge, which has historically been one of the limiting factor for these devices.
This treatment combined with the use of MCPs in low gain mode (gain $<10^4$) can significantly increase the detector lifetime.

The electron cloud produced by the multiplication stage is directed to a pixelated CMOS anode.
This anode is a bare ASIC, namely the Timepix4~\cite{Timepix42020}, that implements the analog and digital front-end electronics.
A detailed description of the Timepix4 main features is given in Section~\ref{sec:timepix}.

The bottom surface of the vacuum tube is formed by a ceramic carrier board providing vacuum feed through connections for the Timepix4 input/output signals. 
This material is chosen to allow good thermal conductivity, necessary to chill the CMOS anode, expected to dissipate $<10~\mathrm{W}$. A heat sink, placed below the ceramic carrier, will allow to maintain the full assembly below $25^{\circ}~\mathrm{C}$.
The ceramic carrier also integrates the Pin Grid Array (\textit{PGA}) connecting the tube to the Data Acquisition (\textit{DAQ}) system, this connection includes both I/O signals and the low-voltage power connections required by the CMOS anode.
We plan to use a custom PGA with $2.54~\mathrm{mm}$ pitch, a  configuration that mimics the standard one used in commercial devices while allowing for the good signal integrity required by the high-speed output of the Timepix4 (up-to $10.556~\mathrm{Gbps}$).
The top surface of the ceramic carrier hosts the pixelated CMOS anode.
This AISC is glued to the carrier and suitable wire bonding pads allow electrical connection to the carrier.

\section{The Timepix4 ASIC}
\label{sec:timepix}

The Timepix4, developed by the Medipix Collaboration, is a $65~\mathrm{nm}$ CMOS ASIC designed for hybrid pixel detectors~\cite{BALLABRIGA201810}.
The ASIC implements a matrix of $512\times448$ pixels distributed over an active area of $6.94~\mathrm{cm^2}$.
A corresponding matrix of bump-pads represent the analog inputs for each pixel; in the described configuration a these bump-pads are used as the anode for each pixel.
The bump-pads are arranged in a matrix with a square pitch of $55~\mathrm{\mu m}$, representing the maximum read-out spatial granularity for the sensor.
The inputs of each pixel are amplified and sent to a time measuring circuitry able to simultaneously measure the Time-of-Arrival (\textit{ToA}) and Time-over-Threshold (\textit{ToT}).
The circuit integrates a Time-to-Digital Converter with $195~\mathrm{ps}$ bin size, equivalent to $56~\mathrm{ps}$ time resolution.

The pixel address together with its ToA and ToT data are encoded in a $66~\mathrm{bit}$ word.
The output words are routed to $16$ high-speed serial links operating at a speed of up to $10.56~\mathrm{Gbps}$ that encode the word with 64b/66b encoding and outputs the data.
The ASIC read-out architecture can be operated in a purely data-driven mode, that is the data is transferred
output only when a pixel is hit with the advantage that the read-out is always active, avoiding the loss of signal and reducing data transfer bandwidth when not need (low rate illumination).
The maximum count rate sustainable by the integrated front end is $3.58\pscalare10^6~\mathrm{hits/mm^2/s}$ corresponding to an average count rate of $10.8~\mathrm{kHz/pixel}$  in case of uniform illumination.

\section{Expected performance}
\label{sec:performance}

The described detector can fully exploit the timing information to improve the temporal and spatial resolution of the system.
Indeed, correction can be applied both on a pixel and pixel-cluster base.

In particular, the ToT information can be used to correct the ToA of each pixel for time-walk effects.
Furthermore, the electron cloud produced by the MCP stack is expected to generate a signal on multiple pixels, that is a cluster.
Reconstruction of cluster using the ToT information can improve the spatial resolution from the $16~\mathrm{\mu m}$, expected for a uniform distribution over a $55~\mathrm{\mu m}$ pitch matrix, to $\approx 5\mathrm{\mu m}$.
A similar improvement in the cluster ToA resolution can be achieved by applying a 3D clustering technique (2D space and time).
In terms, this allows for a multiple sampling of the cluster ToA achieving a timing resolution of few tenths of picoseconds.

\section{Off-detector electronics}
\label{sec:daq}

To read-out the detector, a dedicated off-detector electronic system is under design. 
This system will be composed by two main boards: a detector carrier mezzanine card and a main DAQ board.

The detector carrier mezzanine card has the task of connecting the detector and the main DAQ board.
It hosts the Zero-Insertion-Force socket used to connect the detector to the board.
The input and output signal coming from the detector pass through the socket and they are then buffered and translated so that they can be fed to a Vita57.X family~\cite{vita} connector providing the connection to the main board.
The latter connection is chosen to maintain compatibility with existing read-out systems, like the SPIDR4 board developed by NIKHEF, while being able to exploit existing commercial hardware.
The mezzanine also hosts the voltage regulator grid supplying power to the ASIC.

The main DAQ board will route the high-speed and control signals from a FMC connector directly to a Xilinx Kintex Ultrascale~\cite{kcu} FPGA.
This device will execute minimal data pre-processing (re-ordering, masking, clustering) on the detector output before sending the data to a DAQ server over multiple 10Gbps ethernet connections.
A dedicated 1Gbps ethernet connection to a  control server will be used to control and configure the detector and to monitor the output data.

\section{Conclusion}
\label{sec:conclusion}

We presented the design of a single-photon 'hybrid' detector allowing to detect up to $10^9~\mathrm{photons/s}$ with a simultaneous measurement of time of arrival and position.
This detector will be able to obtain unprecedented resolutions thanks to the excellent timing and position resolution performance of MCPs and of the newly developed 65 nm CMOS pixelated anode, the Timepix4.
These performances go much beyond the state-of-the-art and will open new research paths in different fields of science like
high-energy physics, life sciences and other emerging fields.

\acknowledgments
This project has received funding from the European Research Council (ERC) under the European Union's Horizon 2020 research and innovation programme (Grant agreement No. 819627).

\end{document}